\documentstyle[12pt,twoside]{article}
\begin{document}
\begin{center}

{\Large\bf 
ANISOTROPIC QUANTUM COSMOLOGICAL\\[5PT]
MODELS: A DISCREPANCY BETWEEN\\[5PT]
MANY-WORLDS AND dBB INTERPRETATIONS\\[5PT]}
\medskip
 
{\bf F.G. Alvarenga\footnote{e-mail: flavio@cce.ufes.br}, A.B. Batista\footnote{e-mail: brasil@cce.ufes.br}
J. C. Fabris\footnote{e-mail: fabris@cce.ufes.br}, S.V.B. Gon\c{c}alves\footnote{e-mail: sergio@cce.ufes.br}}\\
  \medskip
Departamento de F\'{\i}sica, Universidade Federal do Esp\'{\i}rito Santo, 
CEP29060-900, Vit\'oria, Esp\'{\i}rito Santo, Brazil \medskip\\

\end{center}
 
\begin{abstract}
In the isotropic quantum cosmological perfect fluid model, the initial singularity can be avoided,
while the classical behaviour is recovered asymptotically.
We verify if initial anisotropies can also be suppressed in a quantum
version of a classical anisotropic model where gravity is coupled to a perfect fluid.
Employing a Bianchi I cosmological model, we obtain a "Schr\"odinger-like" equation where the matter
variables play de role of time. This equation has a hyperbolic signature.
It can be explicitly solved and a wave packet is constructed. The expectation value of the
scale factor, evaluated in the spirit of the many-worlds interpretation, reveals
an isotropic Universe. On the other hand, the bohmian trajectories
indicate the existence of anisotropies. This is an example where the Bohm-de Broglie and the
many-worlds interpretations are not equivalent. It is argued that this inequivalence is due to
the hyperbolic structure of the "Schr\"odinger-like" equation.

\vspace{0.7cm}

PACS number(s): 04.20.Cv., 04.20.Me
\end{abstract}

\section{Introduction}

One of the main hopes regarding quantum cosmology is the possibility to obtain
the initial conditions that will determine the ulterior evolution of the Universe
when its classical regime is reached \cite{halliwell,adm}. For example, it is expected
that the isotropy and homogeneity can be achieved due to the existence of a quantum phase,
prior to the actual classical phase.
Moreover, one expects that the initial singularity problem, which is one of the main drawback of the standard cosmological
scenario, can be circumvented by the presence of quantum effects. In fact, it has already been shown explicitly that
the classical singularity disappears in quantum cosmological scenario where gravity is coupled to a perfect fluid \cite{rubakov,gotay,nivaldo,flavio1}.
In the present work, we address ourselves to another important question concerning the initial conditions of the
Universe: can quantum effects suppress initial anisotropies, leading to an isotropic Universe as soon as
the classical regime is reached? This is not the first time this question is addressed, of course,
(see, for example, \cite{lidsey,gurovich,nelson2}). However, our approach distinguish from the
previous ones by the description employed for the perfect fluid, which allows in principle the identification
of a time coordinate. It permits, as consequence, to compare in detail different interpretations schemes, like
the many-worlds \cite{tipler} and the Bohm-de Broglie (dBB) \cite{holland,nelson} ones in order to obtain specific previsions for the evolution
of the Universe and of the initial anisotropies. We will verify that there is a serious discrepancy between the results
obtained through one scheme and the other, one showing an always isotropic Universe,
while the other one indicates the existence of initial anisotropies which disappear asymptotically,
as it happens in the corresponding classical model. This discrepancy reveals that the quantization of gravity coupled to
perfect fluid leads to a very particular quantum system where some aspects of more ordinary systems,
like the equivalence of the many-worlds and ontological approaches to quantum mechanics, are lost.
\par
It would be perhaps worthfull to
make some general considerations before to describe more in detail the problem to be treated here.
It must be stressed, for example, that the meaning of the Wheeler-DeWitt equation
and of the corresponding wave function of the Universe, which are the basic entities
in quantum cosmology, have no consensus until now.
Besides the technical problems connected with the Wheeler-DeWitt equation, a functional equation
formulated in the superspace, the space of all possible three-dimensional metrics where no explicit
time variable appears\cite{isham}(what constitutes, of course, a serious problem), the interpretation of the results which can be obtained in specific situations is yet an open subject.
Given a specific configuration (gravity plus perfect fluid, for example, as it is the case here),
it is possible in principle to find its corresponding wave function, for example by freezing an infinity
of degrees of freedom. This minisuperspace approach sacrifies of course some important aspects of
the Wheeler-DeWitt equation, like its
functional character.
But, even if the Wheeler-DeWitt
equation may be solved under some restrictive asumptions like the use of a minisuperspace,
it remains the question of how to extract unambiguously predictions from the wave function of the Universe.
\par
In spite of all the controversies, quantum cosmology has experienced some progress in these last years.
As an example, it has been shown by Brown and Kuchar \cite{kuchar} that the coupling of gravity to a dust fluid can
lead to the recovering of the notion of time at quantum level. The Wheeler-DeWitt equation can be recast in a
form similar to that of the Schr\"odinger equation, and a Hilbert space structure can be constructed.
More generally, when a perfect fluid is coupled to gravity, and degrees of freedom are attributed to
the fluid by expressing it through some suitable set of variables, as it happens in the Schutz's formalism for the
description of
perfect fluid \cite{schutz1,schutz2}, a Schr\"odinger-type equation can be obtained, with the fluid variables playing the role of
time. The expectation value of some dynamical quantities can be evaluated, in the spirit of the so-called
many-worlds interpretation,
and a quantum scenario can be sketched. Alternatively, the bohmian trajectories can be computed.
\par
Such approach has obviously some limitations. In the quantum regime, we could expect that fundamental fields
should be considered, instead of a perfect fluid. Moreover, exact solutions can be obtained only in
the minisuperspace, where just some degrees of freedom are kept. But, such quantum perfect fluid models
can be, in spite of their limitations, a very interesting laboratory, in the sense that the role played
by quantum effects in the early Universe can be estimated. In a subject technically difficult and
conceptually so controversial, the possibility of identifying naturally a time coordinate and of obtaining
an explicit behaviour for some dynamical quantities, like the scale factor, may be seen as quite
important achievements, in spite of the limitations of the model for the reasons quoted before.
\par
Many aspects of the isotropic quantum cosmological perfect fluid models in the minisuperspace have been investigated in \cite{rubakov,gotay,nivaldo,flavio1,nelson1,nivaldo1,brasil1}.
In particular, it has been shown that, while the corresponding classical models exhibit an initial singularity, the
quantum scenario is non singular. The results are rigorously the same for the evaluation of
the expectation values and of the bohmian trajectories.
This seems to be a quite relevant result since
it is expected that quantum cosmology can give answer to the question
of what are the initial conditions for the Universe. Here, as it has already been said, we want to answer another
question concerning
the initial conditions of the Universe: to what extent anisotropies that may appear in a classical model can be washed out
by quantum effects.
\par
In some previous work regarding this question, anisotropic models coupled to a scalar field were studied.
In \cite{gurovich,nelson2} a Bianchi I model was considered. In \cite{gurovich}, the emphasis was on the tunneling from a classical
forbiden to a classical allowed region in the configuration space. In \cite{nelson2}
the bohmian trajectories were computed, revealing situations where isotropisation occurs. In \cite{lidsey} all other Bianchi models were taken into
account in view of the evaluation of tunneling process.
But, in all these cases there was no natural time coordinate in the Wheeler-DeWitt equation in
the minisuperspace, and this fact
limits drastically the extraction of unambiguous previsions from the quantum model.
\par
As in the isotropic case, we will consider here an anisotropic model coupled to a perfect fluid, which
is described with the aid of the Schutz's variables, what lead to a natural time coordinate connected
with the fluid variables.
In order to be specific, we will consider a Bianchi I cosmological model.
Through a convenient choice of metric variables, the hamiltonian can be diagonalized. The Wheeler-DeWitt equation
in the minisuperspace can be solved. A wave packet is constructed, satisfying the necessary
boundary conditions.
With this wave packet, and since the notion of time has been recovered by identifying it with the matter variables,
we can try to extract some specific scenario from this model. As it has already been stressed, we can do it essentially in two ways:
by evaluating the expectation value of the scale factor, in the spirit of the many-worlds interpretation
of quantum mechanics, or by computing the bohmian trajectories, in the sense of the ontological
interpretation of quantum mechanics.
However, these two ways of obtaining a prediction (which are in principle connected to different
interpretation's schemes) lead to different results: in the first one, we obtain that the Universe must
be always isotropic; in the second one, anisotropies are present.
\par
Perhaps this unexpected result can be understood by noting that the "Schr\"odinger-like" equation for
the anisotropic perfect fluid cosmological model has a hyperbolic signature in its reduced hamiltonian.
It constitutes, in this sense, a very unusual quantum system for which the equivalence between the
many-worlds and ontological interpretations of quantum mechanics is lost, as it will be discussed more
in detail later.
\par
In the next section, we construct the Wheeler-DeWitt equation for the anisotropic perfect fluid model.
In section 3, we determine the wave function by using the separation of variables method.
In section 4, the expectation value of the scale factor and the bohmian trajectories are obtained. The discrepancy between
them
are settled out. In section 5 we discuss the results and present our conclusions.

\section{Wheeler-DeWitt equation for an anisotropic perfect fluid model}

Our starting point is the action of gravity coupled to a perfect fluid in the Schutz's formalism:
\begin{equation}
\label{action}
S = \int_Md^4x\sqrt{-g}R + 2\int_{\partial M}d^3x\sqrt{h}h_{ab}K^{ab}
+ \int_Md^4x\sqrt{-g}p 
\end{equation}
where $K^{ab}$ is the extrinsic curvature, and $h_{ab}$ is the induced
metric over the three-dimensional spatial hypersurface, which is
the boundary $\partial M$ of the four dimensional manifold $M$; the
factor $16\pi G$ is made equal to one.
The first two terms were first obtained in \cite{adm};
the last term of (\ref{action}) represents the matter contribution to
the total action in the Schutz's formalism for perfect fluids, $p$ being the pressure, which is linked to the
energy density by the equation of state $p = \alpha\rho$. In the
Schutz's formalism \cite{schutz1,schutz2},
the four-velocity is expressed in terms of five potentials $\epsilon$,
$\zeta$, $\beta$, $\theta$ and $S$:
\begin{equation}
U_\nu = \frac{1}{\mu}(\epsilon_{,\nu} + \zeta\beta_{,\nu} +
\theta S_{,\nu})
\end{equation}
where $\mu$ is the specific enthalpy. The variable $S$ is the specific
entropy, while the potentials $\zeta$ and $\beta$ are connected with
rotation and are absent for FRW's type models. The variables $\epsilon$ and
$\theta$ have no clear physical meaning.
The four velocity is subject to the condition
\begin{equation}
U^\nu U_\nu = 1 \quad .
\end{equation}
\par
The metric describing a Bianchi I anisotropic model is given by
\begin{equation}
ds^2 = N^2dt^2 - \biggr(X(t)^2dx^2 + Y(t)^2dy^2 + Z(t)^2dz^2\biggl) \quad .
\end{equation}
In this expression, $N(t)$ is the lapse function.
Using the constraints for the fluid, and after some thermodynamical considerations,
the final reduced action, where surface terms were discarded,
takes the form
\begin{eqnarray}
S = \int dt\biggr[-\frac{2}{N}\biggr(\dot X\dot YZ + \dot X\dot ZY +
\dot Y\dot ZX\biggl) \nonumber\\
+ N^{-1/\alpha}
(XYZ)\frac{\alpha}{(\alpha + 1)^{1/\alpha + 1}}(\dot\epsilon +
\theta\dot S)^{1/\alpha + 1}\exp\biggr(- \frac{S}{\alpha}\biggl)
\biggl] \quad .
\end{eqnarray}
\par
At this point, is more suitable to redefine the metric coefficients as
\begin{equation}
X(t) = e^{\beta_0 + \beta_+ + \sqrt{3}\beta_-}\quad ,\quad Y(t) = e^{\beta_0 + \beta_+ - \sqrt{3}\beta_-}\quad ,
\quad Z(t) = e^{\beta_0 - 2\beta_+} \quad . 
\end{equation}
Using these new variables, the action may be simplified further, leading
to the gravitational lagrangian density
\begin{equation}
L_G = -6\frac{e^{3\beta_0}}{N}\{\dot\beta_0^2 - \dot\beta_+^2 - \dot\beta_-^2\} \quad .
\end{equation}
From this expression, we can evaluate the conjugate momenta:
\begin{equation}
p_0 = - 12\frac{e^{3\beta_0}}{N}\dot\beta_0 \quad , \quad p_+ = 12\frac{e^{3\beta_0}}{N}\dot\beta_+ \quad ,
\quad
p_- = 12\frac{e^{3\beta_0}}{N}\dot\beta_- \quad .
\end{equation}
The matter sector may be recast in a more suitable form through the canonical transformations
\begin{equation}
T = p_Se^{-S}p_\epsilon^{-(\alpha + 1)} \quad , \quad 
\Pi_T = p_\epsilon^{\alpha + 1}e^S \quad , \quad
\bar\epsilon = \epsilon - (\alpha + 1)\frac{p_S}{p_\epsilon} \quad ,
\quad \bar p_\epsilon = p_\epsilon \quad .
\end{equation}
The final expression for the total hamiltonian is
\begin{equation}
\label{hamilton}
H = Ne^{-3\beta_0}\biggr\{- \frac{1}{24}(p_0^2 - p_+^2 - p_-^2) + e^{3(1-\alpha)\beta_0}p_T\biggl\} \quad .
\end{equation}
\par
The lapse function $N$ plays the role of a lagrangian multiplier in (\ref{hamilton}).
It leads to the constraint
\begin{equation}
H = 0 \quad .
\end{equation}
The quantization procedure consists in considering the hamiltonian as
an operator which is applied on a wave function
\begin{equation}
\hat H\psi = 0
\end{equation}
taking at the same time the momenta as operator, in the present case in
the coordinate representation (we use natural units where $\hbar = 1$):
\begin{equation}
\hat p_i = -i\frac{\partial}{\partial\beta_i} \quad .
\end{equation}
Since the momentum associated to the matter degrees of freedom appears linearly in the hamiltonian,
we can identify it with a time coordinate
\begin{equation}
\hat p_T = - i\frac{\partial}{\partial T} \quad .
\end{equation}
Due to the canonical transformations employed before,
this new time is related to the cosmic time $t$ by
$dt = e^{3\alpha\beta_0}dT$.
In this way, we end up with the Wheeler-DeWitt equation, in the minisuperspace, for an anisotropic Universe
filled with a perfect fluid:
\begin{equation}
\label{wdwe}
\biggr(\frac{\partial^2}{\partial\beta^2_0} - \frac{\partial^2}{\partial\beta^2_+} -
\frac{\partial^2}{\partial\beta^2_-}\biggl)\psi = 24ie^{3(1 - \alpha)\beta_0}\frac{\partial\psi}{\partial T} \quad .
\end{equation}

\section{Construction of the wave packet}

Now, our goal is to solve (\ref{wdwe}) and to construct the corresponding wave packet.
To do so, we use the separation of variable method.
First, we write the wave function as
\begin{equation}
\psi(\beta_0,\beta_+,\beta_-,T) = \phi(\beta_0,\beta_+,\beta_-) e^{-iET} \quad ,
\end{equation}
leading to the equation
\begin{equation}
\biggr(\frac{\partial^2}{\partial\beta^2_0} - \frac{\partial^2}{\partial\beta^2_+} -
\frac{\partial^2}{\partial\beta^2_-}\biggl)\phi = 24Ee^{3(1 - \alpha)\beta_0}\phi \quad .
\end{equation}
The function $\phi$ is then
written as
\begin{equation}
\phi(\beta_0,\beta_+,\beta_-) = \Upsilon_0(\beta_0)\Upsilon_+(\beta_+)\Upsilon_-(\beta_-) \quad ,
\end{equation}
leading to the equation
\begin{equation}
\frac{\partial^2_0\Upsilon_0}{\Upsilon_0} + 24Ee^{3(1 - \alpha)\beta_0}
- \frac{\partial^2_+\Upsilon_+}{\Upsilon_+} - \frac{\partial^2_-\Upsilon_-}{\Upsilon_-} = 0
\end{equation}
where we have simplified in a obvious way the notation for the partial derivatives.
The natural ansatz for the functions $\Upsilon_\pm$ is
\begin{equation}
\Upsilon_\pm = C_\pm e^{ik_\pm\beta_\pm} \quad ,
\end{equation}
where $C_\pm$ are constants and $k_\pm$ are the separation parameters. This separation parameters
must be real otherwise the wave function is not normalizable.
\par
The equation determining the behaviour of $\Upsilon_0$ takes then the form,
\begin{equation}
\label{de}
{\Upsilon_0}'' + \biggr(24Ee^{3(1-\alpha)\beta_0} + (k_+^2 + k_-^2)\biggl)\Upsilon_0 = 0 \quad,
\end{equation}
the primes meaning derivatives with respect to $\beta _0$. It is easily to see that the parameter
$E$ must be positive.
The previous equation can be solved through the redefinitions
\begin{equation}
a = e^{\beta_0} \quad , \quad y = a^r \quad , \quad r = \frac{3}{2}(1 - \alpha) \quad .
\end{equation}
after what (\ref{de}) takes the form of a Bessel's equation:
\begin{equation}
{\ddot\Upsilon}_0 + \frac{{\dot\Upsilon}_0}{y} + \biggr(\frac{24E}{r^2} + \frac{k^2}{r^2}\frac{1}{y^2}\biggl)\Upsilon_0 = 0
\end{equation}
where $k^2 = k^2_+ + k^2_-$ and the dots are derivatives with respect to $y$.
The solution is
\begin{equation}
\Upsilon_0 = C_1J_\nu\biggr(\frac{\sqrt{24E}}{r}a^r\biggl) + C_2J_{-\nu}\biggr(\frac{\sqrt{24E}}{r}a^r\biggl) \quad ,
\end{equation}
with $\nu = ik/r$, $C_{1,2}$ being integration constants.
\par
The final expression for the wave function is then
\begin{equation}
\Psi = e^{i(k_+\beta_+ + k_-\beta_-)}\biggr[\bar C_1J_\nu\biggr(\frac{\sqrt{24E}}{r}a^r\biggl) +
\bar C_2J_{-\nu}\biggr(\frac{\sqrt{24E}}{r}a^r\biggl)\biggl]e^{-iET}
\end{equation}
where $\bar C_{1,2}$ are combinations of the preceding integration constants.
We want now to construct a superposition of these solutions, generating a regular wave packet.
In principle, this can be achieved by considering the integration constants as gaussian functions of the
integration parameters $k_\pm$ and $E$. The general case constitutes a hard problem from the
technical point of view. However, since the variables $\beta_+$ and $\beta _-$ appear in a symmetric form
in (\ref{wdwe}), we may consider, for simplicity, the final wave function as independent of one of them, which amounts
to fix one the corrresponding separation parameter $k_+$ or $k_-$ equal to zero. 
From here on we will consider
$k_- = 0$. Notice that the final results would
be the same if we had imposed $k_+ = 0$ and $k_- \neq 0$. Hence, even if the anisotropic models are not
analyzed in all their generality, a large class of them is covered in what follows.
\par
Fixing $k_- = 0$, the wave packet is given by
\begin{equation}
\label{eo}
\Psi = \int e^{ik_+\beta_+}\biggr\{\bar C_1J_\nu\biggr(\frac{\sqrt{24E}}{r}a^r\biggl) +
\bar C_2J_{-\nu}\biggr(\frac{\sqrt{24E}}{r}a^r\biggl)\biggl\}e^{-iET}dk_+dE \quad .
\end{equation}
In principle, in the expression for $\nu$ it appears the modulus of $k_+$ while in the first exponential in (\ref{eo})
we have $- \infty < k_+ < + \infty$.
We will consider a superposition of both Bessel's functions in such a way that the expression
for the wave packet may be written as
\begin{equation}
\Psi = \int_{-\infty}^{+\infty}\int_0^\infty A(k_+,q)e^{ik_+\beta_+}J_\nu\biggr(qa^r\biggl)e^{-iq^2T}dk_+dq \quad ,
\end{equation}
with $q = \frac{\sqrt{24E}}{r}$ and
\begin{equation}
A(k_+,q) = e^{-\gamma k_+^2}q^{\nu + 1}e^{-\lambda q^2} \quad .
\end{equation}
In this case, the integrals can be explicitly calculated, leading to the wave packet
\begin{equation}
\label{wp}
\Psi = \frac{\Psi_0}{B} \exp\biggr[- \frac{a^{2r}}{4B} - \frac{(\beta_+  + C(a,\beta_+))^2}{4\lambda}\biggl]
\end{equation}
where $\Psi_0$ is a constant and
\begin{equation}
B = \lambda + isT \quad , \quad C(a,\beta_+) = \ln a - \frac{2}{3(1 - \alpha)}\ln B \quad , \quad
s = - \frac{3(1 - \alpha)^2}{32} \quad .
\end{equation}
Notice that the wave packet given by (\ref{wp}) is square integrable, and it vanishes in the extremes of the interval of validity of
the variables $a = e^{\beta_0}$ and $\beta_+$, being consequently regular as it is physically required.
Remark that the wave packet (\ref{wp}) is indeed a solution of the equation (\ref{wdwe}), as it can
be explicitly verified.

\section{The scenario for the Universe}

Having the expression for the wave function of the Universe, it is time now to obtain a specificic
prediction for the behaviour of the dynamical functions in this model.
To do so, there is two options: to evaluate the expectation values of the functions describing
the evolution of the scale factors, in the spirit
of the many-worlds interpretation of quantum mechanics; to evaluate the bohmian trajectories
for those functions, in the realm of the ontological interpretation of quantum mechanics.
In all known cases, these different procedures lead to the same results (to an explicit example
treating the isotropic version of the present model, see \cite{nivaldo1}). The former
procedure is possible in our case since we have a time coordinate $T$.
We will evaluate the behaviour of the functions $\beta_0$ and $\beta_+$ using these
two procedures.
\par
Before to do this, let us just recall the classical solutions for the Bianchi I cosmological model
with a barotropic perfect fluid described by $p = \alpha\rho$. For the time parametrization
$dt = a^{3\alpha}dT$, $t$ being the cosmic time, the functions $X$, $Y$, and $Z$ admit the solution
\begin{eqnarray}
X(T) &=& e^{\beta_0 + \beta_+ + \sqrt{3}\beta_-}= X_0\biggr(T + c\biggl)^\frac{1 + 2s_1}{3(1 - \alpha)}\biggr(T - c\biggl)^\frac{1 - 2s_1}{3(1 - \alpha)} \quad ,
\\
Y(T) &=& e^{\beta_0 + \beta_+ - \sqrt{3}\beta_-} = Y_0\biggr(T + c\biggl)^\frac{1 + 2s_2}{3(1 - \alpha)}\biggr(T - c\biggl)^\frac{1 - 2s_2}{3(1 - \alpha)} \quad ,
\\
Z(T) &=& e^{\beta_0 - 2\beta_+} = Z_0\biggr(T + c\biggl)^\frac{1 + 2s_3}{3(1 - \alpha)}\biggr(T - c\biggl)^\frac{1 - 2s_3}{3(1 - \alpha)} \quad ,
\end{eqnarray}
where $c$ is constant, and $s_1$, $s_2$ and $s_3$ are parameters such that
\begin{equation}
s_1 + s_2 + s_3 = 0 \quad , \quad s_1^2 + s_2^2 + s_3^2 = 6 \quad .
\end{equation}
Notice that there is an initial singularity, near which the Universe is very anisotropic, becoming isotropic
asymptotically.
\par
Let us return now to the computation of the quantum scenario through the use of the many-worlds and
ontological interpretations of quantum mechanics.

\subsection{Expectation values of the dynamical variables}

Given the wave function $\Psi$, the expectation value of a variable $\beta_i$ is obtained in the usual way:
\begin{equation}
\label{ev}
<\beta_i> = \frac{\int_{-\infty}^{+\infty}\int_{-\infty}^{+\infty}e^{3(1 - \alpha)\beta_0}\Psi^*\beta_i\Psi d\beta_0d\beta_+}{\int_{-\infty}^{+\infty}\int_{-\infty}^{+\infty}e^{3(1 - \alpha)\beta_0}\Psi^*\Psi d\beta_0d\beta_+} \quad .
\end{equation}
The unusual measure in the integrals is due to the requirement that the reduced hamiltonian in (\ref{wdwe}) must
be symmetric \cite{nivaldo1,brasil1}.
\par
The first step is to calculate the denominator, which will be a common term for
the computation of $<\beta_0>$ and $<\beta_+>$. Using again the definition $a = e^{\beta_0}$ and integrating in
$\beta_+$, we obtain
\begin{equation}
\int_0^\infty\int_{-\infty}^{\infty}a^{2-3\alpha}\Psi^*\Psi dad\beta_+ =
\sqrt{2\gamma\pi}F(T)\int_0^\infty a^{2-3\alpha}\exp{\biggr(-\frac{\lambda a^{3(1-\alpha)}}{2B^*B}\biggl)}da \quad ,
\end{equation}
where
\begin{equation}
F(T) = \frac{\exp{(\frac{C_I^2}{2\gamma})}}{B^*B} \quad ,
\end{equation}
and
\begin{eqnarray}
C(a,\beta_+) &=& C_R + iC_I \quad ,\nonumber\\
C_R = \ln a - \frac{1}{3(1 - \alpha)}\ln 4B^*B \quad &,&
\quad C_I = \frac{- 2}{3(1 - \alpha)}\arctan(\frac{sT}{\lambda})
\quad .
\end{eqnarray}
On the other hand, with $\beta_i = \beta_0$ in (\ref{ev}) we find for the numerator:
\begin{equation}
\int_0^\infty a^{2-3\alpha}\Psi^*\Psi\ln a\;da\;d\beta_+ = \frac{F(T)\sqrt{2\gamma\pi}}{9(1 - \alpha)^2}
\biggr[\frac{2B^*B}{\lambda}\biggl]\biggr\{\ln\biggr(\frac{2B^*B}{\lambda}\biggl)
+ n\biggl\} \quad ,
\end{equation}
where we have noted
\begin{equation}
\quad n
= \int_0^\infty \exp(-u)\ln u\,du \sim - 0.577\quad ,
 \quad u = \frac{\lambda}{2B^*B}a^{3(1 - \alpha)}
\quad .
\end{equation}
Hence,
\begin{equation}
<\beta_0> = \frac{1}{3(1 - \alpha)}\biggr\{\ln\biggr(\frac{2\vert B\vert^2}{\lambda}\biggl) + n\biggl\}
\quad .
\end{equation}
This result leads to
\begin{equation}
e^{<\beta_0>} = (XYZ)^{1/3} = a_0\biggr[1 + \frac{s^2T^2}{\lambda^2}\biggl]^{\frac{1}{3(1 - \alpha)}} \quad ,
\end{equation}
where $a_0$ is a constant. This is the same result as in the isotropic
case \cite{nivaldo1}. Consequently, the space volume evolves as in the corresponding isotropic case.
\par
So, the anisotropies must be represented by the function $\beta_+$, whose expectation value will be computed in what
follows.
We will evaluate now the numerator of (\ref{ev}) with $\beta_i = \beta_+$. Integrating in $\beta_+$ and
expressing $\beta_0$ in terms of $a$ as before, we find:
\begin{eqnarray}
\int_{-\infty}^{+\infty}\int_{-\infty}^{+\infty}e^{3(1 - \alpha)}\Psi^*\beta_+\Psi\;d\beta_0\;d\beta_+ =
- \sqrt{\pi}\biggr\{I_1 - \frac{\ln(4B^*B)}{3(1 - \alpha)}I_2\biggl\}F(T) \quad , \\
I_1 = \int_0^\infty a^{2-3\alpha}\exp\biggr\{- \lambda\frac{a^{3(1 - \alpha)}}{2B^*B}\biggl\}\ln a\;da
\quad , \quad I_2 = \int_0^\infty a^{2-3\alpha}\exp\biggr\{- \lambda\frac{a^{3(1 - \alpha)}}{2B^*B}\biggl\} da
\end{eqnarray}
The integrals $I_1$ and $I_2$ take the form,
\begin{equation}
I_1 = \frac{1}{9(1 - \alpha)^2}\biggr[\frac{2B^*B}{\lambda}\biggl]\biggr[n + \ln\biggr(\frac{2B^*B}{\lambda}\biggl)\biggl] \quad , \quad
I_2 = \frac{1}{3(1 - \alpha)}\frac{2B^*B}{\lambda} \quad .
\end{equation}
Using the previous result for the denominator of (\ref{ev}) we find finally
\begin{equation}
<\beta_+> = \frac{1}{3(1 - \alpha)}\frac{1}{\sqrt{2\gamma}}\biggr\{\ln(2\lambda) - n\biggl\} \quad .
\end{equation}
Surprisingly, the expectation value of $\beta_+$ does not depend on time. Consequently, the predicted result
for the evolution of the Universe in this case is exactly the same as in the isotropic case: there is no
anisotropy during all the evolution of the Universe.

\subsection{Computation of the bohmian trajectories}

The result found in the last section is quite unexpected. There is no trace of the anisotropies existing
in the classical model in the corresponding quantum analysis. We note that in (\ref{wdwe}), the variables
$\beta_+$ and $\beta_-$ appear symmetrically, what it is not the case in the classical model.
But, in order to verify better the meaning of this result, we will evaluate the bohmian trajectories
which determine the behaviour of a quantum system in the ontological interpretation of quantum mechanics.
In principle, the results furnished by the bohmian trajectories must be equivalent to those obtained
through the computation of the expectation values. 
\par
In the ontological interpretation of quantum mechanics, the wave function is written as
\begin{equation}
\label{wf}
\Psi = R\exp(iS) \quad ,
\end{equation}
where $R$ is connected with the amplitude of the wave function, and $S$ to its phase. When (\ref{wf}) is
inserted in the Schr\"odinger's equation, the real and imaginary parts of the resulting expression leads
to the conservation of probability and to a Hamilton-Jacobi's equation supplemented by a term which is identified
as the quantum potential, which leads to the quantum effects distinguishing the quantum trajectories from
the classical ones.
\par
In this formulation of quantum mechanics, the trajectories (which are real trajectories)
corresponding to a dynamical variable $q$ with a conjugate momentum $p_q$, are given by
\begin{equation}
\dot p_q = \frac{\partial S}{\partial q} \quad .
\end{equation}
Hence, the ontological formulation of quantum mechanics
leads to a natural identification of a time coordinate, what is very important for
quantum cosmology where
in general there is no explicit time coordinate.
\par
Let us consider the wave function (\ref{wp}). Putting in the form (\ref{wf}), the phase reads,
\begin{equation}
S(\beta_0,\beta_+,T) = - \arctan\biggr(\frac{sT}{\lambda}\biggl) + \frac{sTa^{3(1 - \alpha)}}{4B^*B} -
\frac{C_I}{2\gamma}(\beta_+ + C_R) \quad ,
\end{equation}
where all quantities are defined as before.
The conjugate momenta associated to the dynamical variables $\beta _0$ and $\beta _+$ read
\begin{equation}
p_0 = - 12a^{2 - 3\alpha}\dot a \quad , \quad p_+ = 12a^{3(1 - \alpha)}\dot\beta_+ \quad ,
\end{equation}
where we have explicitly used the time parametrization such that the lapse function is given
by $N = a^{3\alpha}$.
The bohmian trajectories are then given by the expressions
\begin{eqnarray}
\label{bt1}
- 12a^{2 - 3\alpha}\dot a &=& 3(1 - \alpha)\frac{sT}{4B^*B}a^{3(1 - \alpha)} - \frac{C_I}{2\gamma} \quad , \\
\label{bt2}
12a^{3(1 - \alpha)}\dot\beta_+ &=& - \frac{C_I}{2\gamma} \quad ,
\end{eqnarray}
dots representing derivatives with respect to $T$. Combining (\ref{bt1},\ref{bt2}), we find
\begin{equation}
- 12a^{2 - 3\alpha}\dot a = 3(1 - \alpha)\frac{sT}{4B^*B}a^{3(1 - \alpha)} + 12a^{3(1 - \alpha)}\dot\beta_+ \quad .
\end{equation}
This last equation lead after integration to the expression
\begin{equation}
ae^{\beta_+} = D\biggr[\lambda^2 + s^2T^2\biggl]^\frac{1}{3(1 - \alpha)} \quad .
\end{equation}
Reinserting the relation in the equations (\ref{bt1},\ref{bt2}) we can obtain the following solutions
to $a$ and $\beta_+$:
\begin{eqnarray}
a &=& \biggr(\frac{-1}{24s\lambda\gamma}\biggl)^\frac{1}{3(1 - \alpha)}\biggr[\lambda^2 + s^2T^2\biggl]^\frac{1}{3(1 - \alpha)}\biggr[\arctan^2\biggr(\frac{sT}{\lambda}\biggl) + E\biggl]^\frac{1}{3(1 - \alpha)} \quad , \\
\beta_+ &=& - \frac{1}{3(1 - \alpha)}\ln\biggr\{\arctan^2\biggr(\frac{sT}{\lambda}\biggl) + E\biggl\} + \ln\biggr\{\biggr[-24s\lambda\gamma\biggl]^\frac{1}{3(1- \alpha)}\biggl\} + \ln D
\quad ,
\end{eqnarray}
where $E$ and $D$ are integration constants. Remember that $s < 0$.
\par
In opposition to the expressions obtained for the expectation values of $\beta_0$ (which is connected to $a$)
and $\beta_+$ in the preceding subsection, the bohmian trajectories predict an anisotropic Universe.
Until this point, this strange discrepancy is not so catrastrophic: in order the bohmian trajectories coincide with
the results for the expectation value for some quantity, the integration constants that appear in the
former must be averaged over an initial distribution given by the modulo of the wave function at $T = 0$.
At $T = 0$, we have
\begin{eqnarray}
a(T=0) &=& \biggr(\frac{- \lambda E}{24s\gamma}\biggl)^\frac{1}{3(1 - \alpha)} \quad , \\
\beta_+(T=0) &=& \ln\biggr\{\biggr[\frac{- 24s\lambda\gamma D^{3(1 - \alpha)}}{E}\biggl]^\frac{1}{3(1 - \alpha)}\biggl\} \quad .
\end{eqnarray}
Hence,
\begin{equation}
{\cal R} = \Psi^*\Psi\vert_{T=0} = \frac{\Psi_0^2}{\lambda^2}e^{\frac{E}{48s\gamma} - \frac{1}{18(1 - \alpha)^2\gamma}\ln^2\biggr[\frac{D^{3(1 - \alpha)}}{4}\biggl]} \quad .
\end{equation}
\par
For $a$ and $\beta_+$ the average on the initial conditions leads to the integral expressions
\begin{eqnarray}
\bar a(T) &=& \int_0^\infty\int_0^\infty{\cal R}\;a(T)\;dE\;dD \quad , \\ 
\bar \beta_+(T) &=& \int_0^\infty\int_0^\infty{\cal R}\;\beta_+(T)\;dE\;dD \quad ,
\end{eqnarray}
These expressions can be recast in the following form:
\begin{eqnarray}
\label{i1}
\bar a(T) &=& \beta_2\biggr\{- \frac{\lambda^2 + s^2T^2}{24s\lambda\gamma}\biggl\}^\frac{1}{3(1 - \alpha)}\int_0^\infty
\exp\biggr[\frac{x}{48s\lambda\gamma}\biggl]\biggr[\arctan^2\biggr(\frac{sT}{\lambda}\biggl) + x\biggl]^\frac{1}{3(1 - \alpha)}\;dx \quad , \\ 
\label{i2}
\bar \beta_+ &=& \beta_1 - \frac{\beta_2}{3(1 - \alpha)}\int_0^\infty \exp\biggr[\frac{x}{48s\lambda\gamma}\biggl]\ln\biggr\{\biggr[\arctan^2\biggr(\frac{sT}{\lambda}\biggl) + x\biggl]\biggl\}\;dx \quad ,
\end{eqnarray}
where $\beta_{1,2}$ are numbers given by
\begin{eqnarray}
\beta_1 &=& 48s\bar\Psi_0^2\frac{\gamma}{\lambda}\int_0^\infty \exp\biggr[ - \frac{\ln^2y}{2\gamma}\biggl]\;dy \quad , \\
\beta_2 &=& \frac{\bar\Psi_0^2}{\lambda^2}\int_0^\infty \exp\biggr[ - \frac{\ln^2y}{2\gamma}\biggl]\;dy \quad ,\\
\bar\Psi_0^2 &=& \Psi_0^2\exp\biggr[\frac{1}{2\gamma}\frac{\ln 2}{9(1 - \alpha)^2}\biggl] \quad .
\end{eqnarray}
In the expressions above, we have written $x = E$ and $y = D$. The variables $x$ and $y$ were restricted to positive values in order
to assure that the metric functions are real.
Even if the integrals (\ref{i1},\ref{i2}) seem to admit no simple closed expressions, it is evident that they are time dependent. The
behaviour of $\beta_+$ and $a$ in function of time are displayed in
figures 1 and 2 for $\alpha = 0$ and $\lambda = \gamma = 1$, indicating the presence of anisotropies
even after averaging on the initial conditions, and revealing a non singular scenario, as can be easily
deduced from the above expressions. Notice that these anisotropies disappear in both asymptotes.
Hence, the bohmian trajectories predict an anisotropic Universe even after averaging on the initial
probability distribution, in desagreement with the result obtained through the computation of
the expectation value of the functions $\beta_0$ and $\beta_+$.

\section{Conclusions}

It is generally expected that quantum effects in the very early univere may furnish the set of initial conditions
which will determine the subsequent evolution of the Universe when its classical phase is reached. By initial
conditions we mean here the isotropy and homogeneity. Moreover, it is also expected that those quantum effects
may lead to the avoidance of the initial singularity, one of the major problems of the standard cosmological
model. 
In this work we have tried to analyse the possibility that quantum effects can suppress
initial anisotropies. Specifically, we have studied
a Bianchi I model with a perfect fluid, with an isotropic pressure, employing the Schutz's description for
perfect fluids. This problem has for us two main interests: first, it adds more degrees of freedom
with respect to the isotropic model, since now we have four independent variables instead of just two;
it permits to verify if anisotropies in the early Universe disappear in the quantum model, as it happens
with the initial singularity for the corresponding isotropic one. The employement of Schutz's formalism
for the description of the perfect fluid present in the model allows us to identify quite naturally
a time coordinate associated to the matter degrees of freedom, since the canonical momentum corresponding
to the matter variables appears linearly in the hamiltonian. Hence, the Wheeler-DeWitt equation
can be reduced to a Schr\"ordinger-like equation.
\par
We have solved the Wheeler-DeWitt equation in the minisuperspace. A wave packet was constructed for the special
case that the wave function is independent of one of the variables, namely $\beta_-$. This restriction
was made only because of technical reasons, since it permits to obtain a closed expression for the wave packet.
This wave packet is regular in the sense that it is square integrable and it vanishes in the extreme of
the intervals of $\beta_0$ and $\beta_+$.
Using this wave packet, we have determined the behaviour of the metric functions using first the
many-worlds interpretation of quantum mechanics, which implies to compute the expectation value of those
functions. We found that there is no trace of anisotropies at any moment: the expectation value of
the function $\beta_+$ is constant while the expectation value of $\beta_0$ has essentially the
same expression as in the isotropic version of this problem. All the features of this model are the same as in the isotropic case.
\par
Later, we have determined the behaviour of metric functions employing the ontological interpretation
of quantum mechanics, determining the bohmian trajectories. Surprisingly, in this case the function
$\beta_+$ is no longer a constant, and an initial anisotropic Universe is predicted. Asymptoticaly, it becomes
isotropic like in the classical case. This result is maintained even after the averaging on the initial conditions.
As it is well known \cite{holland,nelson} the bohmian trajectories should lead to the same results that
are obtained computing the expectation values after averaging on the initial conditions. This equivalence
does not occur for the anisotropic Bianchi I cosmological model.
\par
To our knowledge, this is the first case where those interpretation schemes predict different results.
In principle, since the ontological and the many-worlds interpretations of quantum mechanics are
precisely "interpretations" procedures, they must furnish the same results which are "seen" from
different perspective, as we have already stressed. In the present case, the results themselves are different.
What are the reasons for this discrepancy? We guess that the reason for this unexpected feature lies in the fact
that the "Schr\"odinger-like" equation obtained after quantizing the anisotropic perfect
fluid model has a hyperbolic signature in its "spatial" sector. This leads to two different problems: first,
the energy of a free-particle is not positive definite anymore; second, in some sense the functions
$\beta_+$ and $\beta_-$ play also the role of time, in a way similar to what happens in the Klein-Gordon equation.
For both reasons, the equivalence between the ontological and many-worlds approaches to the problem
is broken. This rises the question of which approach to use, a very intringuing problem that it is
not addressed in the present paper.
\par
To verify the hypothesis that the discrepancy found in the results obtained using
the many-worlds and ontological interpretations of quantum mechanic, we change the
signature of the hamiltonian (\ref{hamilton}) by force, making the transformations
$p_+^2 \rightarrow - p_+^2$ and $p_-^2 \rightarrow - p_-^2$. In this way, the
hamiltonian (\ref{hamilton}) gains an elliptical structure.
In this case, discarding the momentum $p_-$ by simplicity, what leads to
two dimensional problem, and performing the canonical transformations
\begin{eqnarray}
x &=& \sqrt{\frac{32}{3}}\frac{e^{\frac{3}{2}(1 - \alpha)\beta_0}}{1 - \alpha} \quad , \quad 
p_x = \frac{1}{24}p_0e^{-\frac{3}{2}(1 - \alpha)\beta_0} \quad , \\
y &=& \sqrt{24}e^{\frac{3}{2}(1 - \alpha)\beta_0}\beta_+ \quad , \quad p_y = \frac{1}{24}p_+e^{-\frac{3}{2}(1 - \alpha)\beta_0} \quad ,
\end{eqnarray}
the hamiltonian (\ref{hamilton}) takes the form
\begin{equation}
H = - p_x^2 - p_y^2 + p_T \quad ,
\end{equation}
which is the hamiltonian for a free-particle in two dimensions. In the free-particle problem, it is
easy to verify that the many-worlds and ontological interpretations give the same results \cite{holland}.
Hence, the discrepancy found above disappears if the the signature of the spatial part of the
hamiltonian is made elliptical.

\vspace{0.5cm}
{\bf Acknowledgements:} We thank CNPq (Brazil) for partial financial support.

\pagebreak

\vspace{2.0cm}

\centerline{\bf Figure captions}

\vspace{5.0cm}

\leftline{Figure 1: Behaviour of $\beta_+$ for $\alpha = 0$, $\lambda = \gamma = 1$}

\vspace{1.0cm}

\leftline{Figure 2: Behviour of $a$ for $\alpha = 0$, $\lambda = \gamma = 1$}
\end{document}